\documentclass{article}
\usepackage{graphics}
\usepackage{epsfig}

\begin{document}

\begin{center} {\large \bf Electromagnetically induced transparency with
degenerate atomic levels}
\end{center}
\bigskip
\begin{center}
\textbf{V. A. Reshetov, I. V. Meleshko}\\
\bigskip
\textit{Department of General and Theoretical Physics, Tolyatti
State University, 14 Belorousskaya Street, 446667 Tolyatti, Russia}
\end{center}

\begin{abstract}
For the coherently driven $\Lambda$-type three-level systems the
general ready-to-calculate expression for the susceptibility tensor
at the frequency of the weak probe field is obtained for the
arbitrary polarization of the strong coupling laser field and
arbitrary values of the angular momenta of resonant atomic levels.
The dependence of the relative difference in the group velocities of
the two polarization components of the probe field on the
polarization of the coupling field is studied.
\end{abstract}

\section{Introduction}

The phenomenon of the electromagnetically induced transparency
(EIT), when the optical properties of the media for the weak probe
field are coherently controlled by the strong coupling field, has
been extensively studied in the recent years and  provided a number
of attractive applications \cite{n1}. The most prominent feature of
EIT is a spectacular reduction of the group velocity of light pulses
in the resonant media \cite{n2,n3}. The strong coupling field with
the definite polarization produces also the optical anisotropy for
the probe field, which reveals itself in the electromagnetically
induced birefringence and polarization rotation of the probe field
\cite{n4,n5}. Among the most promising applications of EIT is the
implementation of quantum memory \cite{n6,n7,n8}. Such memory based
on the controlled  adiabatic deceleration and acceleration of
single-photon pulses in the resonant media was suggested in
\cite{n9,n10}, and soon the possibility of storage of light pulses
was realized in rubidium vapor in the experiment \cite{n11}. The
recent experiments \cite{n12,n13,n14} on EIT-based quantum memory
demonstrate the continuously improving efficiency and fidelity.
There are different ways to encode the single-photon qubit, for
example, in two spectral components of the photon pulse, as it was
recently proposed in \cite{n15}, but the most natural way for qubit
encoding is provided by the photon two polarization degrees of
freedom, as it was implemented in the experiments \cite{n12,n13}. To
store the photon polarization qubit its both polarization components
must be effectively stopped in the medium, however the group
velocities of these two components may differ essentially due to the
optical anisotropy induced by the polarization of the coupling
field. The experiments on EIT are performed on the the three-level
$\Lambda$-type systems with degenerate levels, which are in many
cases the hyperfine structure components of alkali atoms degenerate
in the projections of the atomic total angular momentum on the
quantization axis. The theoretical treatment of EIT on such levels
involves the solution of the equations for the atomic density matrix
and the number of elements of this matrix increases drastically with
the increase of the level angular momentum values. The objective of
the present article is to obtain the general expression for the
linear susceptibility tensor at the frequency of the probe field,
which describes the polarization properties of EIT, and to study the
dependence of the relative difference in the group velocities of the
two polarization components of the probe field on the polarization
of the coupling field. The essential point about this expression is
that the susceptibility tensor may be easily calculated by means of
standard matrix operations for the arbitrary polarization of the
driving field and for the arbitrary values of the level angular
momenta.

\section{Basic equations and relations}

\begin{figure}[t]\center
\includegraphics[width=7cm]{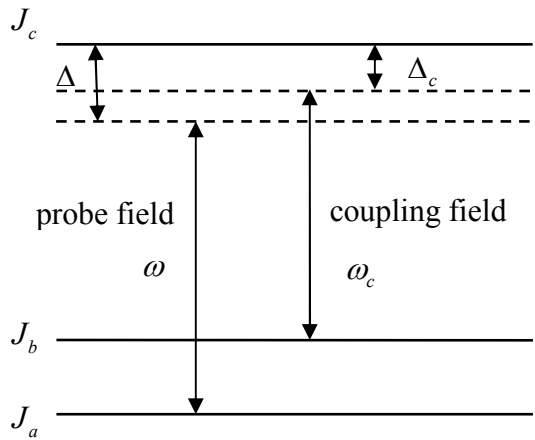}
\caption{The level diagram.}
\end{figure}

We consider the weak probe field propagating along the sample axis
$Z$ with the carrier frequency $\omega$, which is in resonance with
the frequency $\omega_{0}$ of an optically allowed transition
$J_{a}\rightarrow J_{c}$ between the ground state $J_{a}$ and the
excited state $J_{c}$, and the strong coherent coupling field
propagating in the same direction with the carrier frequency
$\omega_{c}$, which is in resonance with the frequency $\omega_{c0}$
of an optically allowed transition $J_{b}\rightarrow J_{c}$ between
the long-lived state $J_{b}$ and the same excited state $J_{c}$
(Fig.1). Here $J_{a}$, $J_{b}$ and $J_{c}$ are the values of the
angular momenta of the levels. The electric field strength of the
coupling field (inside the sample as well as outside) and that of
the probe field incident on the sample border $z=0$ may be put down
as follows:
     \begin{equation}\label{q1}
\textbf{E}_{c}=e_{c}\textbf{l}_{c} e^{-i\omega_{c}(t-z/c)}+ c.c.,
     \end{equation}
     \begin{equation}\label{q2}
\textbf{E}_{0}=\textbf{e}_{0} e^{-i\omega (t-z/c)}+c.c.,
     \end{equation}
where $e_{c}$ and $\textbf{l}_{c}$ are the constant amplitude and
the unit polarization vector of the coupling field, while
$\textbf{e}_{0}$ is the constant vector amplitude of the incident
probe field. At some sample point $z$, where the vector amplitude of
the probe field is $\textbf{e}$, the evolution of the atomic
slowly-varying density matrix $\hat{\rho}$ in the rotating-wave
approximation is described by the equation:
     \begin{equation}\label{q3}
 \frac{d \hat{\rho}}{dt} =
 \frac{i}{2}\left[\hat{V},\hat{\rho}\right] +
 \left(\frac{d\hat{\rho}}{dt}
 \right)_{rel},
     \end{equation}
     \begin{equation}\label{q4}
\hat{V} = 2(\Delta\hat{P}_{c}+\delta\hat{P}_{b}) + \Omega_{c}
(\hat{g}_{c} + \hat{g}_{c}^{\dag}) + \hat{G} + \hat{G}^{\dag},
     \end{equation}
     \begin{equation}\label{q5}
\hat{g}_{c}=\hat{\textbf{g}}_{c}\textbf{l}_{c}^{*},~ \hat{G} =
(2|d|/\hbar)\hat{\textbf{g}}\textbf{e}^{*}.
     \end{equation}
Here $\Delta=\omega-\omega_{0}$ and
$\Delta_{c}=\omega_{c}-\omega_{c0}$ are the frequency detunings from
resonance of the probe and of the coupling fields, while
$\delta=\Delta-\Delta_{c}$, $\hat{P}_{\alpha}$ is the projector on
the subspace of the atomic level $J_{\alpha}$ ($\alpha=a,b,c$),
$\Omega_{c}=2|d_{c}|e_{c}/\hbar$ is the reduced Rabi frequency for
the coupling field, $d=d(J_{a}J_{c})$ and $d_{c}=d(J_{b}J_{c})$
being the reduced matrix elements of the electric dipole moment
operator for the transitions $J_{a}\rightarrow J_{c}$ and
$J_{b}\rightarrow J_{c}$, while $\hat{\textbf{g}}$ and
$\hat{\textbf{g}}_{c}$ are the dimensionless electric dipole moment
operators for the transitions $J_{a}\rightarrow J_{c}$ and
$J_{b}\rightarrow J_{c}$. The matrix elements of the circular
components $\hat{g}_{q}$ and $\hat{g}_{cq}$ ($q=0,\pm 1$) of these
vector operators are expressed through Wigner 3J-symbols \cite{n16}:
    \begin{equation}\label{q6}
  (\hat{g}_{q})^{ac}_{m_{a}m_{c}}=
(-1)^{J_{a}-m_{a}}\left(\matrix{J_{a}&1&J_{c}  \cr
-m_{a}&q&m_{c}}\right),
    \end{equation}
    \begin{equation}\label{q7}
  (\hat{g}_{cq})^{bc}_{m_{b}m_{c}}=
(-1)^{J_{b}-m_{b}}\left(\matrix{J_{b}&1&J_{c}  \cr
-m_{b}&q&m_{c}}\right).
    \end{equation}
Finally, the term $(d\hat{\rho}/dt)_{rel}$ in the equation
(\ref{q3}) describes the irreversible relaxation. Initially the
atoms are at the ground state $a$ the atomic density matrix being
$$\hat{\rho}(0)=\hat{\rho}_{a}.$$
Generally the atoms are at the equilibrium ground state with equally
populated Zeeman sublevels, however they may be prepared at some
special state \cite{n13}. In the linear approximation for the probe
field we obtain from the equations (\ref{q3})-(\ref{q4}) for the
elements of the steady-state ($d\hat{\rho}/dt=0$) atomic density
matrix the following relations:
    \begin{equation}\label{q8}
(\gamma-i\Delta)\hat{\rho}^{ca} = \frac{i}{2}
\left(\Omega_{c}\hat{g}_{c}^{\dag}\hat{\rho}^{ba} +
\hat{G}^{\dag}\hat{\rho}_{a}\right),
    \end{equation}
    \begin{equation}\label{q9}
(\Gamma-i\delta)\hat{\rho}^{ba} = \frac{i}{2}
\Omega_{c}\hat{g}_{c}\hat{\rho}^{ca},
    \end{equation}
where
$$\hat{\rho}^{\alpha\beta}= \hat{P}_{\alpha}\hat{\rho}
\hat{P}_{\beta},~ \alpha,\beta=a,b,c,$$ while the irreversible
relaxation is simply characterized by the two real relaxation rates
-- $\gamma$ for the optically allowed transition $J_{a}\rightarrow
J_{c}$ and $\Gamma$ for the optically forbidden transition
$J_{a}\rightarrow J_{b}$:
$$ \left(\frac{d\hat{\rho}}{dt} \right)_{rel}^{ca} = - \gamma
\hat{\rho}^{ca},~  \left(\frac{d\hat{\rho}}{dt} \right)_{rel}^{ba} =
- \Gamma \hat{\rho}^{ba}.$$ From (\ref{q8})-(\ref{q9}) it follows
immediately:
    \begin{equation}\label{q10}
\hat{\rho}^{ca} =
\frac{i}{2(\gamma-i\Delta)}\hat{U}^{-1}\hat{G}^{\dag}\hat{\rho}_{a},
    \end{equation}
where
    \begin{equation}\label{q11}
\hat{U} = \hat{P}_{c} +
\frac{\Omega_{c}^{2}}{4(\gamma-i\Delta)(\Gamma-i\delta)}
\hat{g}_{c}^{\dag}\hat{g}_{c}.
    \end{equation}
The medium polarization component at point $z$
$$\textbf{P} = \textbf{p} e^{-i\omega t} + c.c. $$
with the frequency $\omega$ of the probe field is expressed through
the density matrix $\hat{\rho}^{ca}$ defined by
(\ref{q10})-(\ref{q11}):
    \begin{equation}\label{q12}
\textbf{p} =
n_{0}|d|Tr_{A}\left\{\hat{\textbf{g}}\hat{\rho}^{ca}\right\},
    \end{equation}
where $n_{0}$ is the concentration of resonant atoms and the trace
is carried out over atomic states. With the two orthonormal vectors
$\textbf{l}_{i}$ in the polarization plane $XY$
($\textbf{l}_{j}\textbf{l}_{k}^{*}=\delta_{jk},~j,k=1,2$) the
equation (\ref{q12}) with an account of (\ref{q10})-(\ref{q11}) and
(\ref{q5}) may be expressed through the susceptibility tensor
$\chi_{jk}$:
    \begin{equation}\label{q13}
p_{j} = \varepsilon_{0}\sum_{k=1}^{2}\chi_{jk}e_{k},
    \end{equation}
where
    \begin{equation}\label{q14}
\chi_{jk}=\frac{i\chi_{0}}{1-i(\Delta/\gamma)}
Tr_{A}\left\{\hat{U}^{-1}\hat{g}_{k}^{\dag}\hat{\rho}_{a}
\hat{g}_{j}\right\},
    \end{equation}
    \begin{equation}\label{q15}
\chi_{0}=\frac{n_{0}|d|^{2}}{\varepsilon_{0}\hbar\gamma},~
\hat{g}_{j}=\hat{\textbf{g}}\textbf{l}_{j}^{*}.
    \end{equation}
With the introduction of the orthonormal set of eigenvectors
$|c_{n}>$ with non-negative eigenvalues $c_{n}^{2}$ of the hermitian
operator $\hat{g}_{c}^{\dag}\hat{g}_{c}$:
$$\hat{g}_{c}^{\dag}\hat{g}_{c}|c_{n}> = c_{n}^{2}|c_{n}>,
n=1,...,2J_{c}+1,$$ the susceptibility tensor (\ref{q14}) may be
transformed to the expression:
    \begin{equation}\label{q16}
\chi_{jk}=\frac{i\chi_{0}}{1-i(\Delta/\gamma)}\sum_{n=1}^{2J_{c}+1}
\frac{<c_{n}|\hat{g}_{k}^{\dag}\hat{\rho}_{a}\hat{g}_{j}|c_{n}>}
{\lambda_{n}},
    \end{equation}
    \begin{equation}\label{q17}
\lambda_{n} = 1 +
\frac{c_{n}^{2}\Omega_{c}^{2}}{4(\gamma-i\Delta)(\Gamma-i\delta)}.
    \end{equation}
Now let $\chi_{k}$ be the two eigenvalues and
    \begin{equation}\label{q18}
\textbf{v}_{k}=\sum_{j=1}^{2}v_{jk}\textbf{l}_{j},~ k=1,2,
    \end{equation}
be the corresponding two eigenvectors of the $2\times 2$
susceptibility matrix $\hat{\chi}=\{\chi_{jk}\}$:
$$\hat{\chi}\textbf{v}_{k}=\chi_{k}\textbf{v}_{k},~ k=1,2.$$
Then, the vector amplitude $\textbf{e}_{0}$ of the incident probe
field after the passage of the distance $z$ in the medium is
transformed to
    \begin{equation}\label{q19}
\textbf{e}=\hat{S}\textbf{e}_{0},~
\hat{S}=\hat{v}\hat{T}\hat{v}^{-1},
    \end{equation}
where $\hat{v}=\{v_{jk}\}$ is the $2\times 2$-matrix, defined by the
equation (\ref{q18}), while the diagonal matrix $\hat{T}$ is
determined by the eigenvalues $\chi_{k}$ of the susceptibility
matrix:
    \begin{equation}\label{q20}
T_{jk}=\delta_{jk}e^{i(\omega/c)zn_{j}},~n_{j}=\sqrt{1+\chi_{j}},
    \end{equation}
as it may be obtained in a usual way from the Maxwell equation
$$\frac{\partial^{2}\textbf{E}}{\partial z^{2}} -
\frac{1}{c^{2}}\frac{\partial^{2}\textbf{E}}{\partial t^{2}} =
\frac{1}{\varepsilon_{0}c^{2}}
\frac{\partial^{2}\textbf{P}}{\partial t^{2}}$$ for the electric
field strength $\textbf{E}$ of the probe field. The intensity $I$
and the $2\times 2$ polarization matrix
$\hat{\sigma}=\{\sigma_{jk}\}$ of the probe field after the passage
of the distance $z$ in the medium are related to the intensity
$I_{0}$ and the polarization matrix $\hat{\sigma}_{0}$ of the
incident probe field by the equations:
    \begin{equation}\label{q21}
\frac{I}{I_{0}}=tr\left(\hat{S}\hat{\sigma}_{0}\hat{S}^{\dag}\right),~
\hat{\sigma}=\frac{\hat{S}\hat{\sigma}_{0}\hat{S}^{\dag}}
{tr\left(\hat{S}\hat{\sigma}_{0}\hat{S}^{\dag}\right)},
    \end{equation}
where transformation matrix $\hat{S}$ is defined by
(\ref{q19})-(\ref{q20}).

 \begin{figure}[t]\center
\includegraphics[width=7cm]{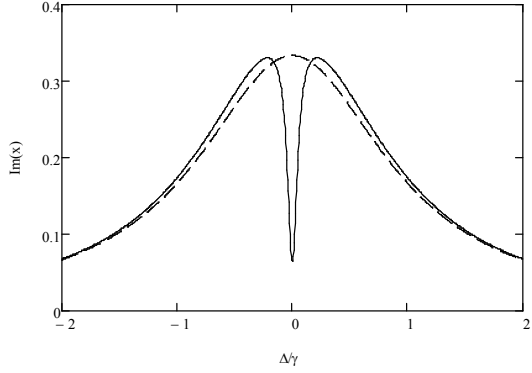}
\caption{The imaginary part of susceptibility
$Im(\chi_{xx}/\chi_{0})$ (dashed line) and $Im(\chi_{yy}/\chi_{0})$
(solid line) versus dimensionless frequency detuning $\Delta/\gamma$
on the transitions with $J_{a}=0$, $J_{b}=J_{c}=1$ at
$\Omega_{c}/\gamma=1$, $\Gamma/\gamma=0.01$ and
$\textbf{l}_{c}=\textbf{l}_{x}$, $\Delta_{c}=0$.}
    \end{figure}

The electromagnetically induced transparency is associated with the
significant reduction of the group velocity of the probe field due
to the steep dispersion at the transparency window. Since the
dispersion is different for the two polarization components of the
probe field, these two components will be differently slowed down in
the medium. Let us calculate the relative difference of the group
velocities for these two polarization components. For the
observation of the electromagnetically induced transparency the
coupling field must be strong enough:
$$\Omega_{c}^{2}\gg \Gamma\gamma.$$
Then, close to the two-photon resonance $\delta\leq \Gamma$ the
general formula (\ref{q16})-(\ref{q17}) for the susceptibility
tensor is simplified to the following expression:
    \begin{equation}\label{q22}
\chi_{jk}=\frac{4\chi_{0}\gamma}{\Omega_{c}^{2}}
\left(\delta+i\Gamma\right)a_{jk},
    \end{equation}
where
    \begin{equation}\label{q23}
a_{jk}=\sum_{n}
\frac{<c_{n}|\hat{g}_{k}^{\dag}\hat{\rho}_{a}\hat{g}_{j}|c_{n}>}
{c_{n}^{2}},
    \end{equation}
is a hermitian $2\times 2$ matrix. The summation in (\ref{q23}) is
carried out over all eigenvectors $|c_{n}>$ of the operator
$\hat{g}_{c}^{\dag}\hat{g}_{c}$ with non-zero eigenvalues
$c_{n}^{2}>0$. The two eigenvectors $\textbf{v}_{k}$ (\ref{q18}) of
the susceptibility matrix $\{\chi_{jk}\}$ coincide with the two
orthonormal eigenvectors of the hermitian matrix $\{a_{jk}\}$, while
its eigenvalues
$$\chi_{k}=\frac{4\chi_{0}\gamma}{\Omega_{c}^{2}}
\left(\delta+i\Gamma\right)a_{k},$$ are expressed through the real
eigenvalues $a_{k}$ of matrix $\{a_{jk}\}$. The group velocities
$V^{gr}_{k}$ of the two probe field components polarized along
eigenvectors $\textbf{v}_{k}$ are determined by the refraction
indices $n'_{k}=Re\left(\sqrt{1+\chi_{k}}\right)$:
$$V^{gr}_{k}=\frac{c}{n'_{k}+\omega (dn'_{k}/d\omega)}.$$
Since $|\chi_{k}|\ll 1$, in the case of steep dispersion
$$\chi_{0}\gamma\omega\gg \Omega_{c}^{2},$$ we obtain for the group
velocities:
    \begin{equation}\label{q24}
V^{gr}_{k}=\frac{c}{n^{gr}_{k}},~ n^{gr}_{k}=
\frac{2\chi_{0}\gamma\omega a_{k}}{\Omega_{c}^{2}}.
    \end{equation}
The relative difference of these two group velocities
    \begin{equation}\label{q25}
\varepsilon=\frac{V^{gr}_{2}-V^{gr}_{1}}{V^{gr}_{2}} =
1-\frac{a_{2}}{a_{1}}
    \end{equation}
is determined by the difference of the two eigenvalues $a_{k}$ of
matrix $\{a_{jk}\}$, $a_{1}$ being the largest of them and the group
velocity $V^{gr}_{1}$ of the probe field component polarized along
the corresponding eigenvector $\textbf{v}_{1}$ being the  smallest
of two.

\section{Discussion}

\begin{figure}[t]\center
\includegraphics[width=7cm]{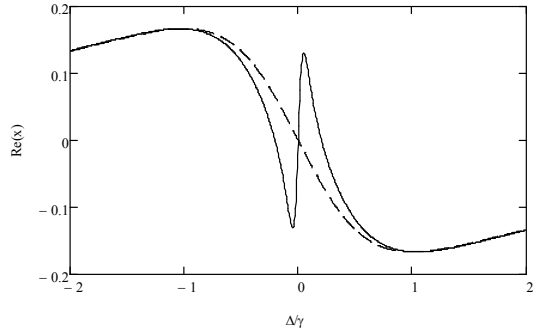}
\caption{The real part of susceptibility $Re(\chi_{xx}/\chi_{0})$
(dashed line) and $Re(\chi_{yy}/\chi_{0})$ (solid line) versus
dimensionless frequency detuning $\Delta/\gamma$ on the transitions
with $J_{a}=0$, $J_{b}=J_{c}=1$ at $\Omega_{c}/\gamma=1$,
$\Gamma/\gamma=0.01$ and $\textbf{l}_{c}=\textbf{l}_{x}$,
$\Delta_{c}=0$.}
   \end{figure}

The $(2J_{b}+1)\times(2J_{c}+1)$ matrices $\hat{g}_{c}$ and
$(2J_{a}+1)\times(2J_{c}+1)$ matrices $\hat{g}_{j}$ ($j=1,2$) of the
electric dipole moment operators may be easily calculated for any
reasonable values of the angular momenta $J_{a}$, $J_{b}$ and
$J_{c}$ of the resonant atomic levels using the known formulae for
Wigner 3J-symbols, and further calculations of the susceptibility
tensor $\hat{\chi}$ (\ref{q14}) and of the transformation matrix
$\hat{S}$ (\ref{q19}), involving standard matrix operations, are
rather simple. In the following numerical calculations we shall
assume the coupling field to be rather strong: $\Omega_{c}/\gamma=1$
and exactly resonant $\Delta_{c}=0$, and the relaxation rate of the
forbidden transition to be rather small: $\Gamma/\gamma=0.01$, as it
is the case for the electromagnetically induced transparency, while
the atoms are initially at the equilibrium state:
$$\hat{\rho}_{a}=\frac{\hat{P}_{a}}{2J_{a}+1}.$$
We shall also consider the linearly polarized coupling field, then
without loss of generality its polarization vector may be directed
along the Cartesian axis $X$: $\textbf{l}_{c}$ = $\textbf{l}_{x}$,
and we shall choose the Cartesian basis
$\textbf{l}_{1}=\textbf{l}_{x}$, $\textbf{l}_{2=}\textbf{l}_{y}$ in
the polarization plane $XY$ for the calculation of the elements of
the susceptibility tensor. With this choice of the Cartesian basis
the polarization matrix $\hat{\sigma}$ (\ref{q21}) of the probe
field may be expressed through the Stokes parameters $\xi_{n}$
($n=1,2,3$):
$$\sigma = \frac{1}{2} \left( \matrix{1+\xi_{3} &\xi_{1}-i\xi_{2}
\cr \xi_{1} +i\xi_{2}&1-\xi_{3}}\right),$$
$$P=\sqrt{\xi_{1}^{2}+\xi_{2}^{2}+\xi_{3}^{2}}$$
being the total degree of polarization,
$$ P_{l}=\sqrt{\xi_{1}^{2}+\xi_{3}^{2}}$$
being the degree of linear polarization, and
$$P_{c}=|\xi_{2}|$$
being the degree of circular polarization.

  \begin{figure}[t]\center
\includegraphics[width=7cm]{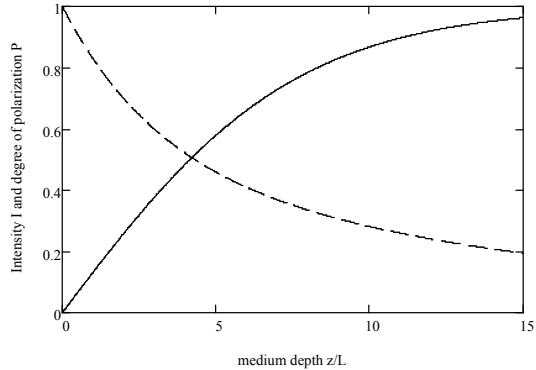}
\caption{The relative intensity $I/I_{0}$ (dashed line) and the
degree of polarization $P$ (solid line) of the initially unpolarized
probe field versus dimensionless medium depth $z/L$
($L=1/k\chi_{0}$) on the transitions with $J_{a}=0$, $J_{b}=J_{c}=1$
at $\Omega_{c}/\gamma=1$, $\Gamma/\gamma=0.01$ and
$\textbf{l}_{c}=\textbf{l}_{x}$, $\Delta=\Delta_{c}=0$.}
  \end{figure}

Let us consider the transitions with the low values of angular
momenta $J_{a}=0$, $J_{b}=J_{c}=1$, which may be realized in
thallium vapor for example. With the linear polarization
$\textbf{l}_{c}$ = $\textbf{l}_{x}$ of the coupling field the
susceptibility tensor becomes diagonal in the Cartesian basis. The
dependencies of the imaginary $Im(\chi_{xx}/\chi_{0})$,
$Im(\chi_{yy}/\chi_{0})$ and real $Re(\chi_{xx}/\chi_{0})$,
$Re(\chi_{yy}/\chi_{0})$ parts of the susceptibility eigenvalues
versus the dimensionless frequency detuning $\Delta/\gamma$,
obtained from the calculations according to the formula (\ref{q16}),
are presented in the Figures 2 and 3. As it follows from these
dependencies the strong linearly polarized (along the axis $X$)
coupling field opens the "transparency window" (solid line) for the
polarization component of the probe field linearly polarized in the
perpendicular direction (along the axis $Y$), while for the
component of the probe field collinearly polarized with the coupling
field (along the axis $X$) this "window" remains shut (dashed line),
which gives rise to rather strong dichroism for the resonant
($\Delta=0$) probe field. This happens because the $X$-polarized
component of the probe field interacts with a single substate
$$|c_{x}>=\frac{1}{\sqrt{2}}\left(|m_{c}=-1>-|m_{c}=1>\right)$$
of the excited level $c$, which is not affected by the coupling
field:
$$\hat{g}_{c}|c_{x}>=0.$$
For the initially unpolarized probe field:
$$\hat{\sigma_{0}}=\frac{1}{2} \left( \matrix{1&0
\cr 0&1}\right),$$ the dependencies of its relative intensity
$I/I_{0}$ (dashed line) and the degree of polarization $P$ (solid
line) versus the dimensionless medium depth $z/L$, where
$L=1/k\chi_{0}$, are presented in the Figure 4. As it may be seen in
this figure, the probe field becomes almost fully polarized with the
degree of polarization $P=0.96$ (linearly along the axis $Y$), while
its intensity is reduced by the factor of 0.19.

\begin{figure}[t]\center
\includegraphics[width=7cm]{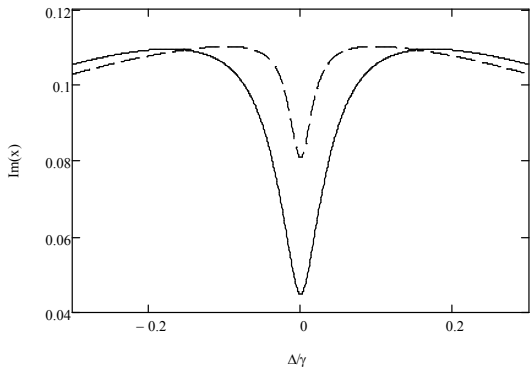}
\caption{The imaginary part of susceptibility
$Im(\chi_{xx}/\chi_{0})$ (dashed line) and $Im(\chi_{yy}/\chi_{0})$
(solid line) versus dimensionless frequency detuning $\Delta/\gamma$
on the transitions with $J_{a}=1$, $J_{b}=J_{c}=2$ at
$\Omega_{c}/\gamma=1$, $\Gamma/\gamma=0.01$ and
$\textbf{l}_{c}=\textbf{l}_{x}$, $\Delta_{c}=0$.}
  \end{figure}

Let us now consider the transitions with larger values of angular
momenta $J_{a}=1$, $J_{b}=J_{c}=2$, which were employed in the
experiment \cite{n4} performed in $^{87}Rb$ vapor. For the linear
polarization $\textbf{l}_{c}$ = $\textbf{l}_{x}$ of the coupling
field the dependencies of the imaginary $Im(\chi_{xx}/\chi_{0})$,
$Im(\chi_{yy}/\chi_{0})$ and real $Re(\chi_{xx}/\chi_{0})$,
$Re(\chi_{yy}/\chi_{0})$ parts of the susceptibility eigenvalues
versus the dimensionless frequency detuning $\Delta/\gamma$ are
presented in the Figures 5 and 6. In this case the "transparency
windows" are open for both polarization components of the probe
field, however, that for the $Y$-component (solid line),
perpendicular to the coupling field, is larger than for the
$X$-component (dashed line), collinear with the coupling field. In
the case of large values of the angular momenta the dichroism is
less than in the case of small ones, as it may be seen in the Figure
7. When the degree of polarization of the initially unpolarized
probe field attains the value of 0.5 its intensity is reduced by the
factor of 0.17.

  \begin{figure}[t]\center
\includegraphics[width=7cm]{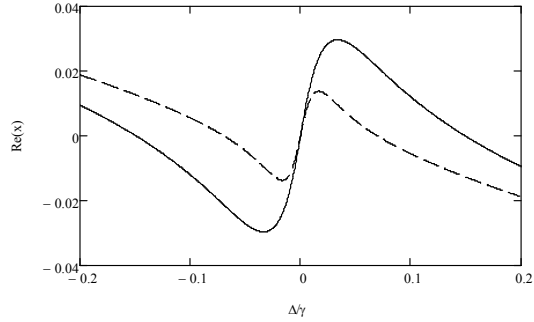}
\caption{The real part of susceptibility $Re(\chi_{xx}/\chi_{0})$
(dashed line) and $Re(\chi_{yy}/\chi_{0})$ (solid line) versus
dimensionless frequency detuning $\Delta/\gamma$ on the transitions
with $J_{a}=1$, $J_{b}=J_{c}=2$ at $\Omega_{c}/\gamma=1$,
$\Gamma/\gamma=0.01$ and $\textbf{l}_{c}=\textbf{l}_{x}$,
$\Delta_{c}=0$.}
  \end{figure}
  \begin{figure}[t]\center
\includegraphics[width=7cm]{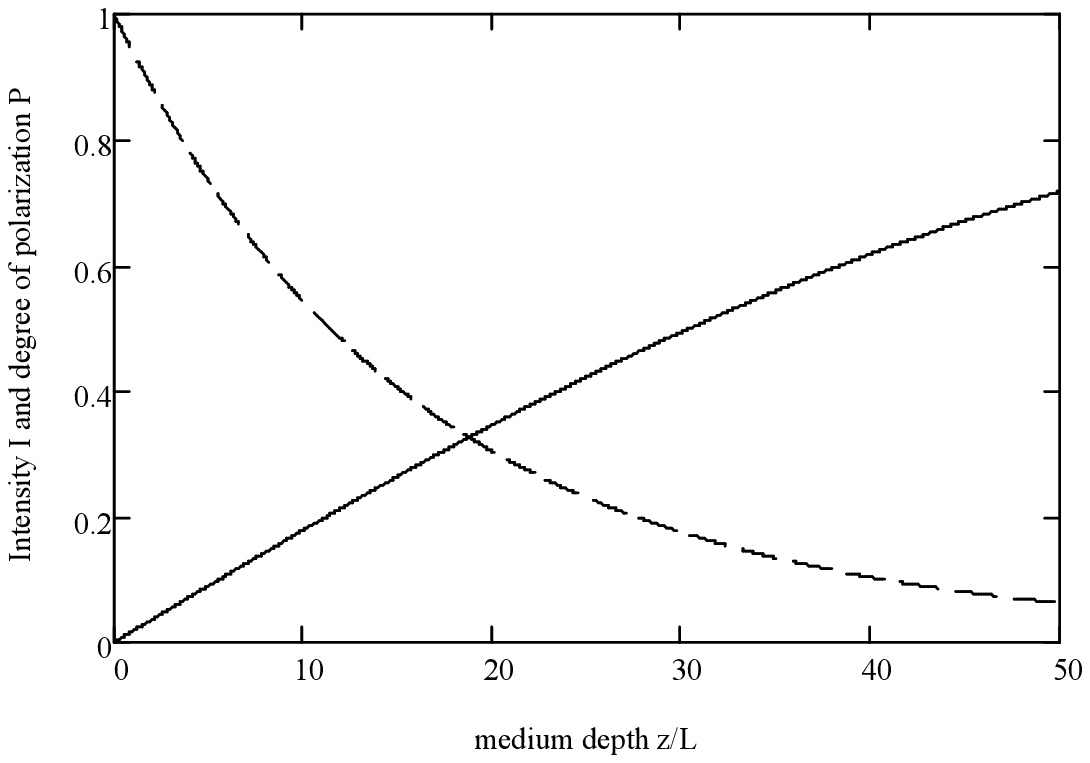}
\caption{The relative intensity $I/I_{0}$ (dashed line) and the
degree of polarization $P$ (solid line) of the initially unpolarized
probe field versus dimensionless medium depth $z/L$
($L=1/k\chi_{0}$) on the transitions with $J_{a}=1$, $J_{b}=J_{c}=2$
at $\Omega_{c}/\gamma=1$, $\Gamma/\gamma=0.01$ and
$\textbf{l}_{c}=\textbf{l}_{x}$, $\Delta=\Delta_{c}=0$.}
  \end{figure}
  \begin{figure}[t]\center
\includegraphics[width=7cm]{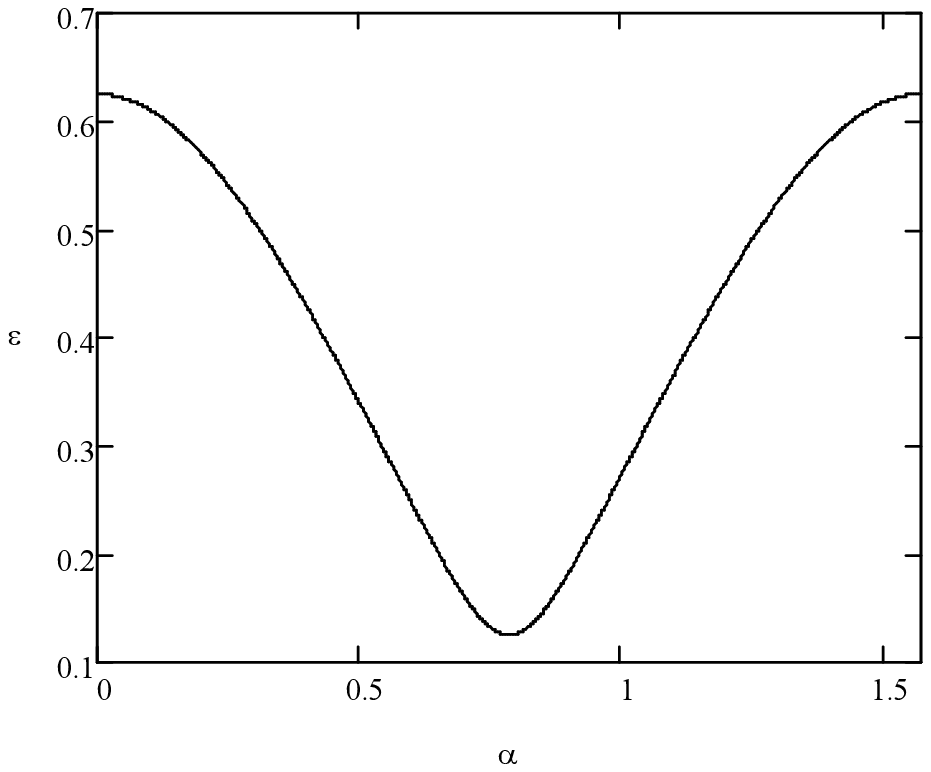}
\caption{The dependence of relative difference
$\varepsilon=(V^{gr}_{2}-V^{gr}_{1})/V^{gr}_{2}$ in the group
velocities of the two polarization components of the probe field on
the ellipticity parameter $\alpha$ of the coupling field on the
transitions with $J_{a}=J_{c}=1$, $J_{b}=2$.}
  \end{figure}

The relative difference $\varepsilon$ in the group velocities of the
two polarization components of the probe field (\ref{q25}) depends
only on the values of total angular momenta of resonant levels, on
the initial atomic state and on the polarization of the coupling
field. Let us consider the coupling field with arbitrary elliptic
polarization in the plane $XY$:
$$l_{cq}=\cos\alpha\delta_{q,-1}-\sin\alpha\delta_{q,1},$$
where parameter $\alpha$ defines the ratio of the lengthes $a_{y}$
and $a_{x}$ of the axes of polarization ellipse along the axes $Y$
and $X$ according to the relation
$$\left|\tan\left(\alpha-\frac{\pi}{4}\right)\right| =
\frac{a_{y}}{a_{x}},$$ while the sign of $\tan(\alpha-\pi/4)$
determines the direction of rotation of the pulse electric field
vector in the plane $XY$. For the atoms at initially equilibrium
state and with the fixed values of the level angular momenta the
relative difference in the group velocities $\varepsilon$ depends
only on the ellipticity parameter $\alpha$ of the coupling field. An
example of such dependence for the levels with the angular momenta
$J_{a}=J_{c}=1$, $J_{b}=2$, employed in the experiment \cite{n13},
is presented in the Figure 8. The maximum difference in the group
velocities $\varepsilon=0.625$ is obtained with the circularly
polarized coupling field ($\alpha=0,\pi/2$), while the minimum
difference $\varepsilon=0.125$ is obtained with the linearly
polarized coupling field ($\alpha=\pi/4$). In the case of circularly
polarized coupling field the polarization component with the larger
group velocity is circularly polarized in the same direction as the
coupling field, while the polarization component with the smaller
group velocity is circularly polarized in the direction opposite to
that of the coupling field. In the case of linearly polarized
coupling field the polarization component with the smaller group
velocity is linearly polarized in same direction as the coupling
field, while the polarization component with the larger group
velocity is linearly polarized in the direction perpendicular to
that of the coupling field. In the experiment \cite{n13} the
coupling field was linearly polarized in the direction of
propagation of the probe field: $l_{cq}=\delta_{q,0}$
($\pi$-polarized), while it propagated in the perpendicular
direction, and the atoms were prepared at the pure Zeeman state
$|J_{a}=1,m_{a}=0>$ with the zero projection on the quantization
axis. Under such conditions we obtain for the tensor (\ref{q23}):
$a_{jk}=1.667\delta_{jk}$, so that the group velocity of the probe
pulse in this case does not depend on its polarization. For the
unprepared atoms, which are initially in the equilibrium state with
equally populated Zeeman sublevels, we obtain
$a_{jk}=0.972\delta_{jk}$, so that in this case the group velocity
of the probe pulse also does not depend on its polarization.

\section{Conclusions}

In the present article the  $\Lambda$-type three-level systems with
degenerate levels  coherently driven by the strong laser field are
considered. The general expression for the susceptibility tensor at
the frequency of the weak probe field is obtained for the arbitrary
polarization of the strong coupling laser field and arbitrary values
of the  momenta $J_{a}$, $J_{b}$, $J_{c}$ of resonant atomic levels.
The numerical calculations, based on this expression, may be easily
performed by means of standard matrix operations for any reasonable
values of angular momenta. Sample calculations for the transitions
with $J_{a}=0$, $J_{b}=J_{c}=1$ and $J_{a}=1$, $J_{b}=J_{c}=2$,
which are employed in the experiments with atomic vapors, were
carried out. It was shown, that the depth of the "transparency
window" for the resonant probe field depends on its polarization
providing rather strong dichroism. Such dependence is greater for
the low values of angular momenta $J_{a}=0$, $J_{b}=J_{c}=1$ and it
is muffled for larger values $J_{a}=1$, $J_{b}=J_{c}=2$. The general
expression for the relative difference in the group velocities of
the two polarization components of the probe field is obtained. This
difference depends on the values of total angular momenta of
resonant levels, on the initial atomic state and on the polarization
of the coupling field. Such dependence on the ellipticity parameter
of the coupling field for the transitions with $J_{a}=J_{c}=1$,
$J_{b}=2$, studied numerically, revealed rather a wide range of
variation of this relative difference -- from 0.625 with the
circularly polarized coupling field to 0.125 with the linearly
polarized one.

{\bf Acknowledgements}

Authors are indebted for financial support of this work to Russian
Ministry of Science and Education (grant 2.2407.2011).

\end{document}